\documentclass[aps,twocolumn,amsmath,amssymb,showpacs,showkeys]{revtex4-1}
\usepackage{graphicx}
\usepackage{epsfig}
\usepackage{dcolumn}
\usepackage{bm}

\newcommand{\la}{\left\langle}         
\newcommand{\ra}{\right\rangle}        
\newcommand{\eq}[1]{Eq.~(\ref{#1})}    
\newcommand{\rf}[1]{Ref.~\cite{#1}}    
\newcommand{\fg}[1]{Fig.~\ref{#1}}     

\begin{document}
\title{Wealth distribution: To be or not to be a Gamma?}

\author{Mehdi Lallouache}
  \email{mehdi.lallouache@ens-cachan.fr}
  \affiliation{D\'epartement de Physique, \'Ecole Normale Sup\'erieure de Cachan, 94230 Cachan, France}
  \affiliation{Chaire de Finance Quantitative, Laboratoire de Math\'ematiques Appliqu\'ees aux Syst\`emes, \'Ecole Centrale Paris, 92290 Ch\^atenay-Malabry, France}

\author{Aymen Jedidi}
 \email{aymen.jedidi@ecp.fr}
  \affiliation{Chaire de Finance Quantitative, Laboratoire de Math\'ematiques Appliqu\'ees aux Syst\`emes, \'Ecole Centrale Paris, 92290 Ch\^atenay-Malabry, France}

\author{Anirban Chakraborti}
  \email{anirban.chakraborti@ecp.fr}
   \affiliation{Chaire de Finance Quantitative, Laboratoire de Math\'ematiques Appliqu\'ees aux Syst\`emes, \'Ecole Centrale Paris, 92290 Ch\^atenay-Malabry, France}

\date{\today}

\begin{abstract}
We review some aspects, especially those we can tackle analytically, of a minimal model of closed economy analogous to the kinetic theory model of ideal gases where the agents exchange wealth amongst themselves such that the total wealth is conserved, and each individual agent saves a fraction ($0\le\lambda\le1$) of
wealth before transaction. We are interested in the special case where the
fraction $\lambda$ is constant for all the agents (global saving propensity) in the closed system. We show by moment calculations that the resulting wealth distribution cannot be the Gamma distribution that was
conjectured in Phys. Rev. E 70, 016104 (2004). We also derive a form for the distribution at low wealth, which is a new result.
\end{abstract}


\maketitle

\section{Introduction}
The distribution of wealth or income in society has been of great interest for many years.
As first noticed by Pareto in the 1890's \cite{Pareto}, the wealth distribution
seems to follow a ``natural law'' where the tail of the distribution is described
by a power-law $f(x)\sim x^{-(1+\alpha)}$. Away from the tail, the distribution is better described by a Gamma or Log-normal distribution known as Gibrat's law \cite{Gibrat}. Considerable investigation with real data during the last ten years revealed that the power-law tail exhibits a remarkable spatial and temporal stability and the Pareto index $\alpha$
is found to have a value between $1$ and $2$ \cite{Dragulescu2001,Aoyama2003}. Even after 110 years the origin of the power-law tail remained unexplained but recent interest of physicists and mathematicians
in econophysics has led to a new insight into this problem (see Refs. \cite{Yakovenko2009,Arnab2007,Chakrabarti2010}).\\
\indent Our general aim is to study a many-agent statistical model
of closed economy (analogous to the kinetic theory model of ideal gases) \cite{Bennati,Angle,Ispolatov1998,Dragulescu2000a,Chakraborti2000a,Chatterjee2003a},
where $N$ agents exchange a quantity $x$, that
may be defined as wealth.
The states of agents are characterized by the wealth
$\{x_i\},~i=1,2,\dots,N$, and the total wealth $W=\sum_{i} x_i$ is conserved.
The evolution of the system is then carried out according to a prescription,
which defines the trading rule between agents.
These many-agent statistical models have $N$ basic units $\{1, 2,\dots, N\}$,
interacting with each other through a pair-wise interaction characterized
by a saving parameter $\lambda$, with $0 \le \lambda \le 1$. We define the equilibrium distribution of wealth $f(x)$ as follows : $f(x) dx $ is the probability that in the steady state of the system, a randomly chosen agent will be found to have wealth between $x$ and $x + dx$.
In these models, if $\lambda$ is equal for all the units, $f(x)$ is fitted quite well by a Gamma-distribution \cite{Anirban2008,Patriarca2004a,Patriarca2004b}
\begin{equation}
\label{gamman}
  f(x) = \frac{1}{\Gamma(n)} \left ( \frac{n}{\la x \ra}\right )^n  x^{n-1}
  \exp \left ( - \frac{nx}{\la x \ra} \right ) ,
\end{equation}
where
\begin{eqnarray}
  \label{D1}
  n
  = \frac{D(\lambda)}{2}
  = 1 + \frac{3\lambda}{1-\lambda}.
  \label{n}
\end{eqnarray}
This equilibrium distribution \eqref{gamman} had been suggested by an analogy with the kinetic theory of gases in $D(\lambda)$ dimensions \cite{Anirban2008,Patriarca2004a,Patriarca2004b}.

In this paper we show by the method of moment calculations that the resulting wealth distribution cannot be the Gamma distribution that was conjectured in \rf{Patriarca2004a,Patriarca2004b}. We also derive the functional form of an upper bound on $f(x)$ at very small $x$.

\section{Many-agent model of a closed economy}
We study many-agent statistical models
of closed economy (analogous to the kinetic theory model of ideal gases),
where $N$ agents exchange wealth $x$.
The states of agents are characterized by the wealth
$\{x_i\},~i=1,2,\dots,N$, and the total wealth $W=\sum_{i} x_i$ is conserved.
The evolution of the system is then carried out according to a prescription,
which defines the trading rule between agents.
At every time step two agents $i$ and $j$ are extracted randomly
and an amount of wealth $\Delta x$ is exchanged between them,
\begin{align}
  x_i' &= x_i - \Delta x \, ,
  \nonumber \\
  x_j' &= x_j + \Delta x \, .
  \label{basic0}
\end{align}
It can be noticed that in this way, the quantity $x$ is conserved
during the single transactions: $x_i'+x_j' = x_i + x_j$
(see \fg{fig:exchange}),
where $x_i'$ and $x_j'$ are the agent wealths
after the transaction has taken place.
Several simple models dealing with different transaction rules have been studied (see the reviews \cite{Yakovenko2009,Arnab2007,Patriarca2010} and references therein). Here we will present a few examples.

\subsection{Basic model without saving: Boltzmann distribution}
\label{sec:basic}
In the first version of the model, the wealth difference $\Delta x$
is assumed to have a constant value
\cite{Bennati},
\begin{equation}
  \Delta x = \Delta x_0 \ .
\end{equation}
This rule, together with the constraint
that transactions can take place only
if $x_i'>0$ and $x_j'>0$, provides a Boltzmann distribution,
see the curve for $\lambda=0$ in \fg{fig:gamma}.
Alternatively, $\Delta x$ can be a random fraction of the wealth
of one of the two agents,
\begin{equation}
  \Delta x = \epsilon x_i
  ~~ {\rm or} ~~
  \Delta x = \epsilon    x_j \ ,
\end{equation}
where $\epsilon$ is a random number uniformly distributed between 0 and 1.
A trading rule based on the random
redistribution of the sum of the wealths of the two agents
had been introduced by Dragulescu and Yakovenko \cite{Dragulescu2000a},
\begin{align}
  x_i' &= \epsilon (x_i + x_j) \, ,
  \nonumber \\
  x_j' &= (1-\epsilon) (x_i + x_j) \, .
  \label{basic1}
\end{align}
Equations (\ref{basic1}) are easily shown to correspond to the trading rule (\ref{basic0}), with
\begin{equation}
  \Delta x =  (1-\epsilon) x_i - \epsilon x_j   \ .
\end{equation}
All the versions of the basic model lead to
an equilibrium Boltzmann distribution, given by
\begin{equation}
  f(x) = \frac{1}{ \la x \ra }\exp\left(-\frac{x}{\la x \ra}\right) \, ,
  \label{BD}
\end{equation}
where the effective temperature of the system is just
the average wealth $\la x \ra$
\cite{Bennati,Dragulescu2000a}.
The result (\ref{BD}) is found to be robust;
it is largely independent of various factors.
Namely, it is obtained for the various forms of $\Delta x$
mentioned above, for a pair-wise as well as multi-agent interactions,
for arbitrary initial conditions \cite{Chakraborti2000a},
and finally, for random or consecutive extraction of the interacting
agents. For the trading rule \eqref{basic1} one can show the convergence towards the Boltzmann distribution
through different methods: Boltzmann equation, entropy maximization, distributional equation, etc.

\begin{figure}
\begin{center}
	\includegraphics[width=\linewidth]
        {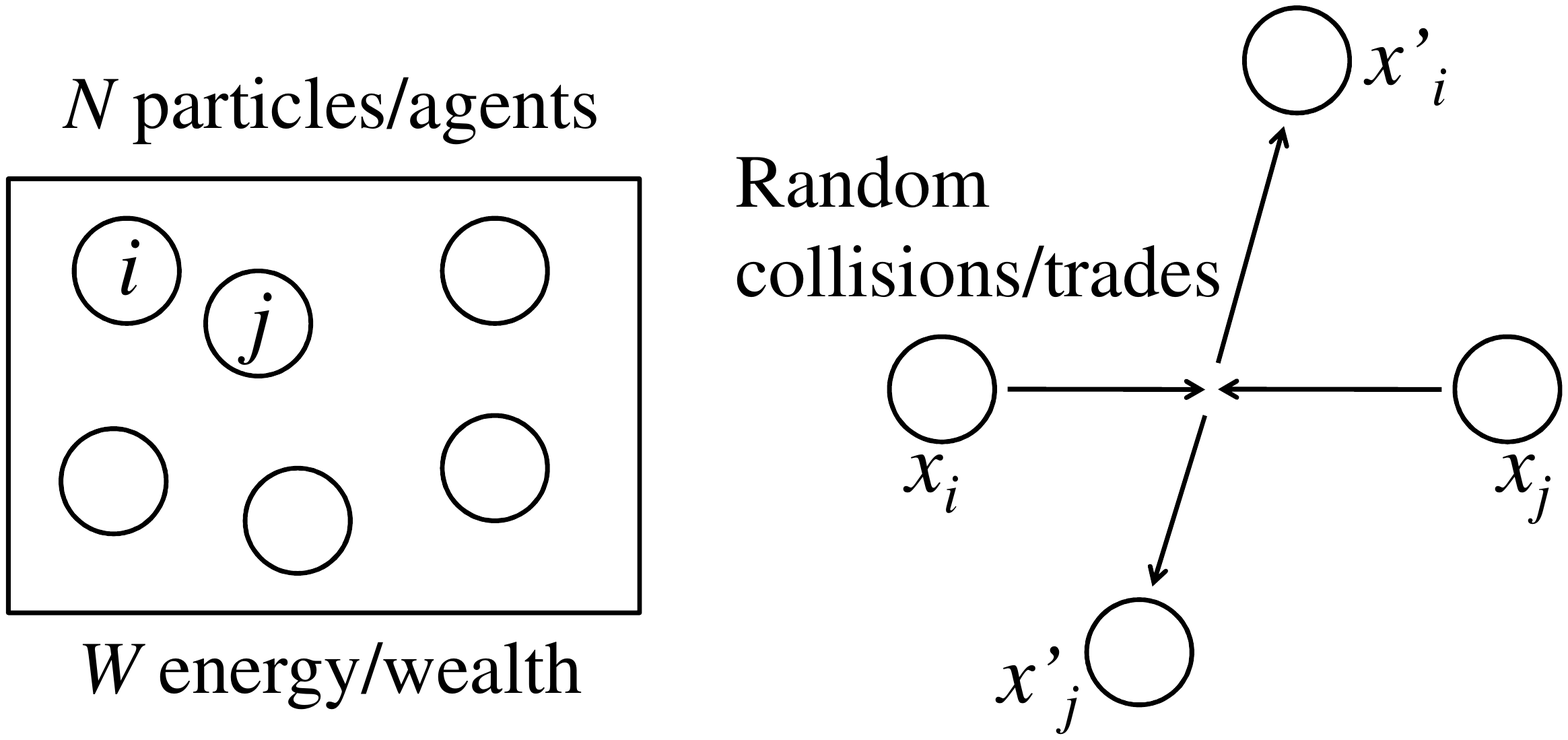}
    \end{center}
\caption{Analogy of the minimal economic model with a classical isolated system
of ideal gas, where the particles are randomly undergoing ``Elastic'' collisions, and exchanging kinetic energy. In the closed economy, the economic agents randomly trade with each other according to some rule and exchange wealth.}
\label{fig:exchange}
\end{figure}
\begin{figure}
  \begin{center}
    \includegraphics[width=\linewidth]{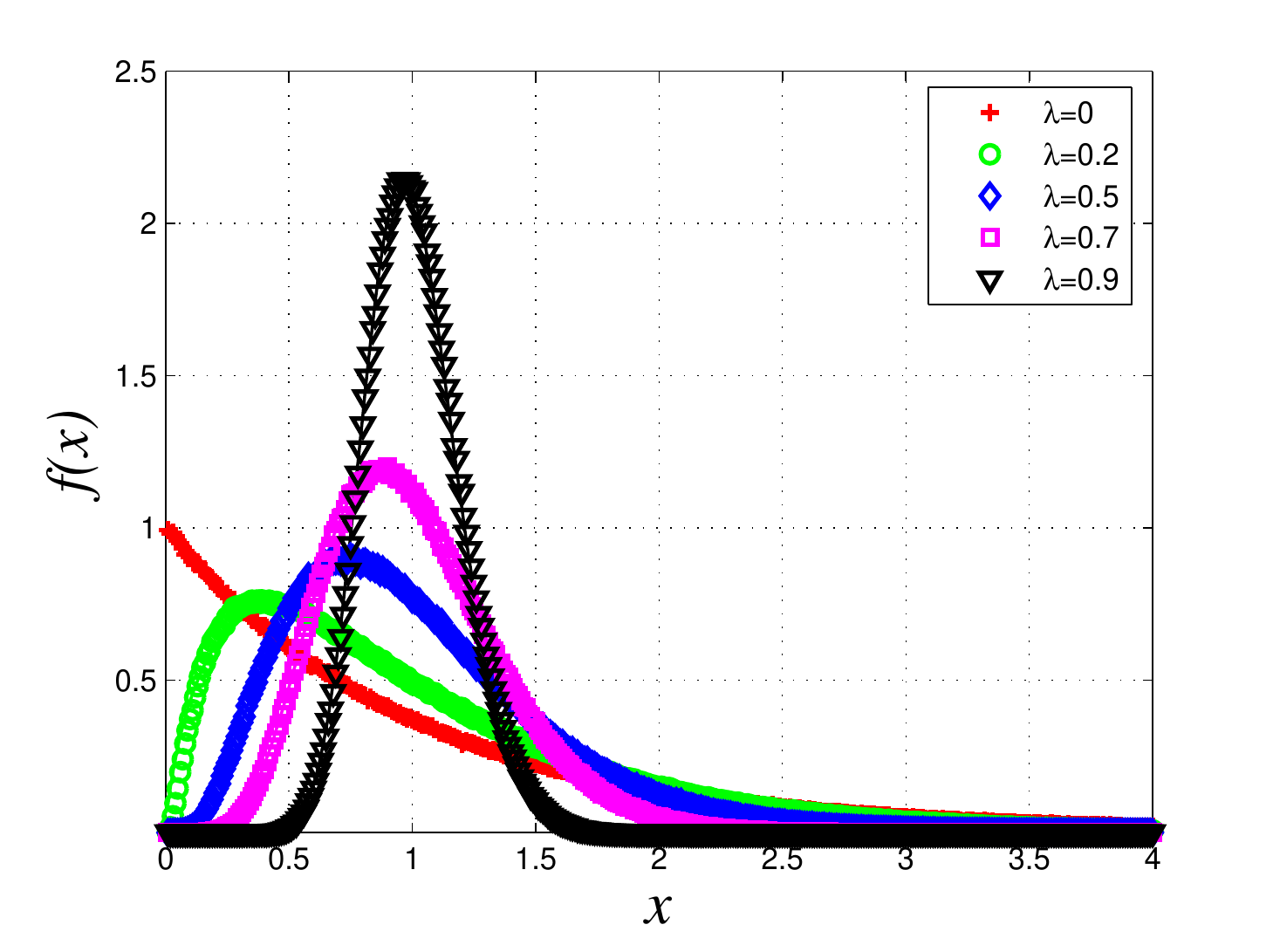}\\
    \includegraphics[width=\linewidth]{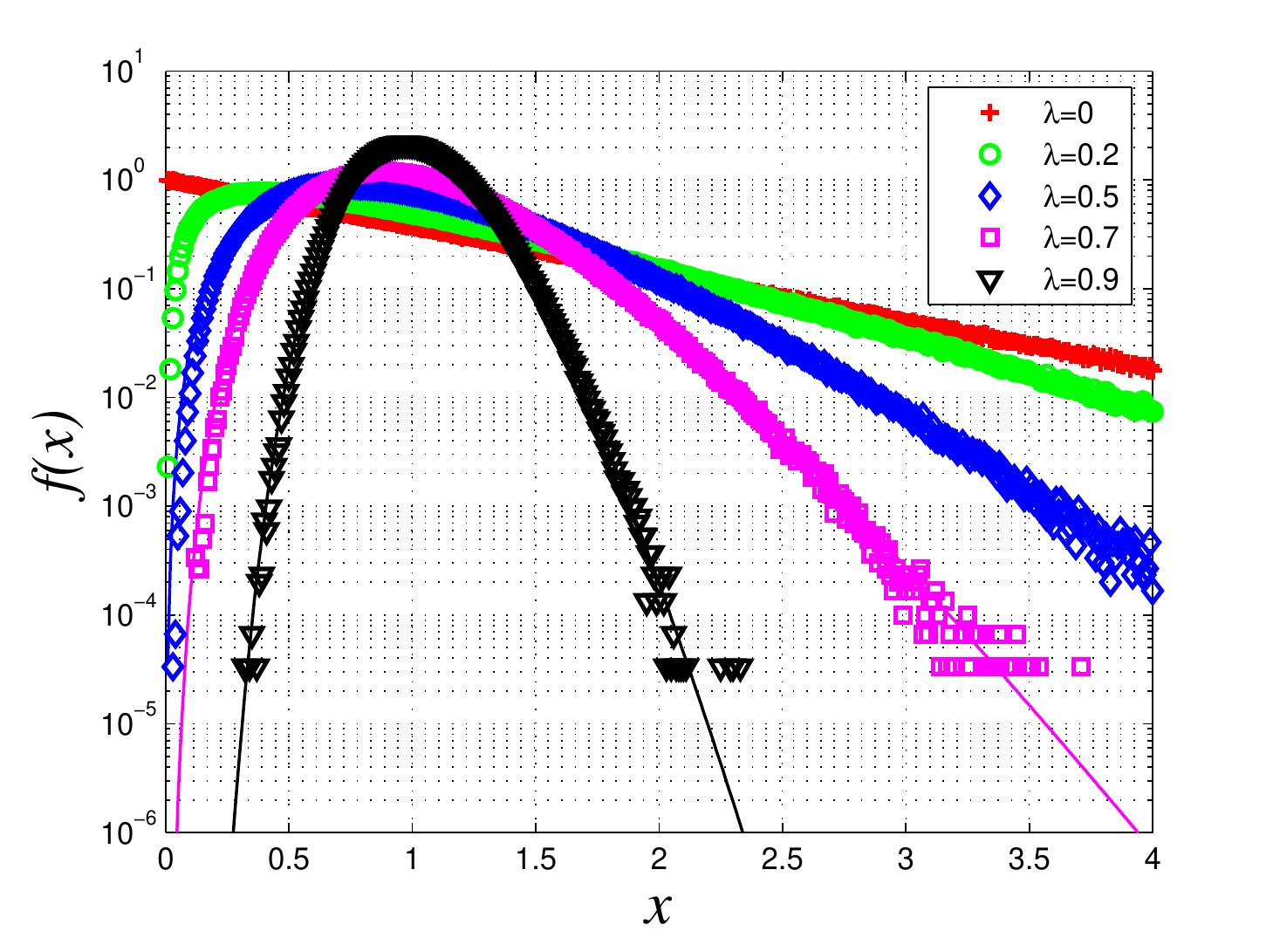}
    \caption{ Probability density for wealth $x$.
    The curve for $\lambda=0$ is the Boltzmann function
    $f(x)=\la x \ra^{-1}\exp(-x/\!\la x \ra)$ for the basic model
    of Sec. \ref{sec:basic}.
    The other curves correspond to a global saving propensity
    $\lambda>0$ (see Sec. \ref{global}).}
    \label{fig:gamma}
  \end{center}
\end{figure}

\subsection{Model with global saving propensity $\lambda$}
\label{global}
A step toward generalizing the basic model and making it more realistic, is
the introduction of a saving criterion regulating the trading dynamics.
This can be practically achieved by defining a saving propensity
$0 \le \lambda \le 1$,
which represents the fraction of wealth which is saved
-- and not reshuffled -- during a transaction.
The dynamics of the model
is as follows
\cite{Chakraborti2000a,Chakraborti2002a}:
\begin{align}
  x_i' &= \lambda x_i + \epsilon (1-\lambda) (x_i + x_j) \, ,
  \nonumber \\
  x_j' &= \lambda x_j + (1-\epsilon) (1-\lambda) (x_i + x_j) \, ,
  \label{sp1}
\end{align}
corresponding to a $\Delta x$ in \eq{basic0} given by
\begin{equation}
  \Delta x =     (1 - \lambda) [ (1-\epsilon) x_i - \epsilon x_j ]  \, .
\end{equation}
This model leads to a qualitatively different equilibrium distribution.
In particular, it has a mode $x_m>0$ and a zero limit for small $x$, see \fg{fig:gamma}. Later we will derive a form
for an upper bound on $f(x)$ at low range.
The functional form of such a distribution was conjectured to be a $\Gamma$-distribution, as given by
Eq.~(\ref{gamman}) on the basis of an analogy with the kinetic theory of gases. Indeed, it is easy to show, starting from the Maxwell-Boltzmann distribution for the particle velocity in a $D$ dimensional gas, that the equilibrium kinetic energy distribution coincides with the Gamma-distribution \eqref{gamman} with $n=\frac{D}{2}$. This conjecture is remarkably consistent with the fitting provided to numerical data \cite{Anirban2008,Patriarca2004a,Patriarca2004b}. In the following section we will show by two different approaches that the conjecture \eqref{gamman} cannot be the actual equilibrium distribution.

\section{Analytical results for model with saving propensity $\lambda$}
\label{analytical}

\subsection{Fixed-point distribution}
\begin{figure}
  \begin{center}
    \includegraphics[width=\linewidth]{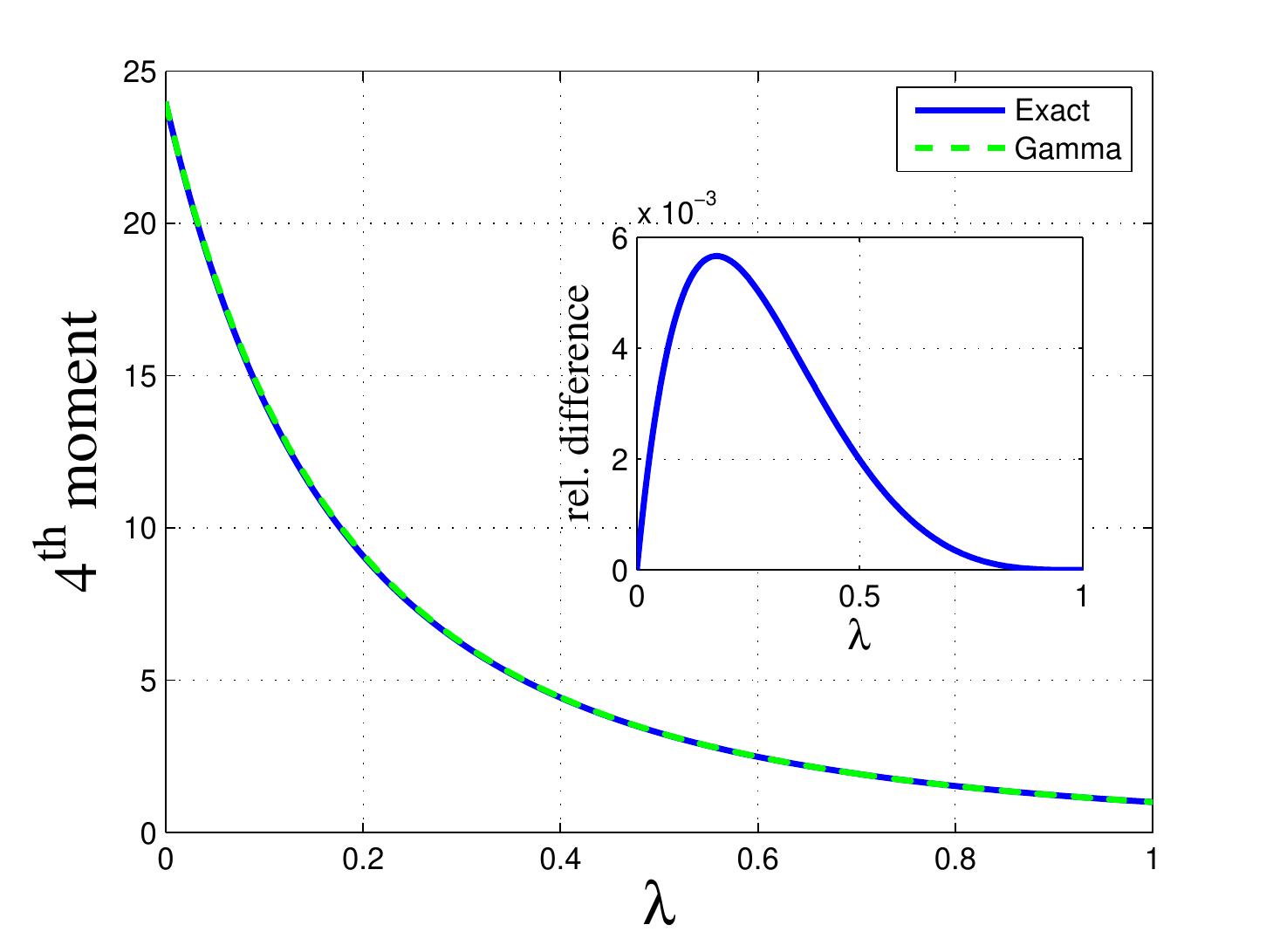}
    \caption{Exact fourth moment eq.\eqref{m1} and Gamma distribution fourth moment eq.\eqref{m2} against $\lambda$. The inset shows the relative difference between exact fourth moment eq.\eqref{m1} and Gamma distribution fourth moment eq.\eqref{m2} against $\lambda$.}
    \label{fourth}
  \end{center}
\end{figure}

Let $X$ be a random variable which stands for the wealth of one agent, at equilibrium and in the limit of an infinite number of agents, we can say from Eq. \eqref{sp1} that the law of $X$ $f$, is a fixed-point distribution of the equation
\begin{equation}
    X \stackrel{\text{d}}{=} \lambda X_1 + \epsilon(1-\lambda)(X_1+X_2),
    \label{distrib}
\end{equation}
where $ \stackrel{\text{d}}{=}$ means identity in distribution and one assumes that the random variables
 $X_1,X_2$ and $X$ have the same probability law, while the variables $X_1,X_2$ and $\epsilon$ are stochastically independent.
It seems difficult to find the distribution of $X$, however, one can compute the moments of $f$. Indeed with \eqref{distrib}, one can write immediately
  \begin{equation}
  \forall m \in \mathbb{N}, \la X^m \ra = \la (\lambda X_1 + \varepsilon(1-\lambda)(X_1+X_2))^m \ra ,
  \label{moments}
  \end{equation}
 and by developing \eqref{moments} one can find the recursive relation
  \begin{equation}
  \la X^m \ra = \sum_{k=0}^{m} \binom{m}{k} \frac{\lambda^{m-k} (1-\lambda)^k}{k+1} \sum_{p=0}^{k} \binom{k}{p}
  \la X^{m-p} \ra \la X^p \ra .
  \label{rec}
  \end{equation}
Using \eqref{rec} with initial conditions $\la X^0 \ra=1$ (normalization) and $\la X^1 \ra=1$ (without loss of generality), we obtain
\begin{align}
\la X^2 \ra &= \frac{\lambda +2}{1+2\lambda} \,,\\
\la X^3 \ra &= \frac{3(\lambda +2)}{(1+2\lambda )^{2}} \,,\\
\la X^4 \ra &= \frac{72+12\lambda -2\lambda ^{2}+9\lambda ^{3}-\lambda ^{5}}{(1+2\lambda )^{2}(3+6\lambda -\lambda ^{2}+2\lambda ^{3})} \label{m1}\, .
\end{align}
Now let us compare theses moments with conjecture \eqref{gamman}'s moments.
Setting $\la x \ra = 1$ in Eq. \eqref{gamman} it is easy to show
\begin{equation}
\la x^k \ra = \frac{(n+k-1)(n+k-2)...(n+1)}{n^{k-1}} \,.
\label{gammamoments}
\end{equation}
Writing \eqref{gammamoments} for $k=2,3,4$ and choosing $n$ as in \eqref{n} we find
\begin{align}
 \la x^2 \ra &= \frac{n+1}{n} = \frac{\lambda +2}{1+2\lambda}  \,, \\
 \la x^3 \ra &= \frac{(n+2)(n+1)}{n^2} = \frac{3(\lambda +2)}{(1+2\lambda )^{2}}  \,, \\
 \la x^4 \ra &= \frac{(n+3)(n+2)(n+1)}{n^3} =  \frac{3(\lambda +2)(4-\lambda)}{(1+2\lambda )^{3}} \label{m2} \,.
 \end{align}
The fourths moments (eqs.\eqref{m1} and \eqref{m2}) are different so the conjecture that the Gamma distribution is an equilibrium solution of this model is wrong. Nevertheless the first three moments coincide exactly which shows that the Gamma-distribution is strangely a very good approximation. Moreover the deviation in the fourth moment is very small (see Fig. \ref{fourth}, which shows that the two curves can hardly be distinguished by the naked eye).
Finding a function that would coincide to higher moments is still an open challenge.
These results are consistent with the ones found by Repetowicz et al. \cite{Repetowicz2005} which will be presented in the following section.

\subsection{Laplace transform analysis}
In this section we will confirm the previous result with a different approach
based on the Boltzmann equation and along the lines of Bassetti et al. \cite{Matthes2008}.
Given a fixed number of $N$ agents in a system, which are allowed to
trade, the interaction rules describe a stochastic
process of the vector variable $(x_1(\tau), \dots, x_N(\tau))$ in discrete
time $\tau$. Processes of this type are thoroughly studied e.g. in the
context of kinetic theory of ideal gases. Indeed, if the variables
$x_i$ are interpreted as energies corresponding to the $i$-th particle,
one can map the process to the mean-field limit of the Maxwell model
of particles undergoing random elastic collisions. The full information about
the process in time $\tau$ is contained in the $N$-particle joint
probability distribution $P_N(\tau, x_1, x_2, \dots, x_N)$. However,
one can write a kinetic equation for one-marginal distribution
function
\[
P_1(\tau,x) = \int P_N(\tau, x, x_2, \dots, x_N)dx_2\cdots dx_N ,
 \]
involving only one- and two-particle distribution functions
\begin{multline}
P_1(\tau+1,x)- P_1(\tau,x) = \bigg \langle \frac{1}{N} \Big[ \int P_2(\tau, x_i, x_j) \notag\\
\Big( \delta( x -\lambda x - (1-\lambda)\epsilon(x_i + x_j)) \notag \\
+ \delta( x - \lambda x - (1-\lambda)(1-\epsilon)(x_i + x_j)) \Big) dx_i  dx_j\\
- 2P_1(\tau, x)\Big] \bigg \rangle,
\end{multline}
which may be continued to give eventually an infinite hierarchy of
equations of BBGKY (Born, Bogoliubov, Green, Kirkwood, Yvon) type \cite{Plischke}. The standard approximation,
which neglects the correlations between the wealth of the agents
induced by the trade gives the factorization
 \[
P_2(\tau, x_i, x_j) =P_1(\tau,x_i)P_1(\tau,x_j),
 \]
which implies a closure of the hierarchy at the lowest level. In
fact, this approximation becomes exact in the thermodynamic limit ($N\to\infty$).
Therefore, the one-particle distribution
function bears all information. Rescaling the time as $t = \frac{2\tau}{N}$
in the thermodynamic limit $N\to\infty$, one obtains for the
one-particle distribution function $f(t,x)$ the
Boltzmann-type kinetic equation
\begin{multline}
\frac{\partial f (t, x)}{\partial t}  = \frac{1}{2} \Big \langle \int f(t, x_i)f(t, x_j)\\
\Big ( \delta( x -\lambda x - (1-\lambda)\epsilon(x_i + x_j))\\
 + \delta( x -\lambda x - (1-\lambda)(1-\epsilon)(x_i + x_j)) \Big ) dx_i \, dx_j\Big \rangle \\
 - f(t,x).
\end{multline}
This equation can be written (see Matthes et al. \cite{Matthes2008}) as
$$\frac{\partial f (t, x)}{\partial t}=Q(f,f)\,,$$ where $Q$ is a \textit{collision operator}.
A collision operator is bilinear and satisfies, for all smooth functions $\phi(x)$
\begin{multline}
 \int_{0}^{\infty} Q(f,f) \phi(x)dx \\
 = \frac{1}{2}  \Big \langle \int_{0}^{\infty} \int_{0}^{\infty} (\phi(x_i')+\phi(x_j')-\phi(x_i)-\phi(x_j)) \\
  f(x_i)f(x_j)dx_i dx_j \Big \rangle,
\end{multline}
where $x_i'$ and $x_j'$ are the post-trade wealth.
With this property the equation can be written in the weak form, for all smooth functions $\phi(x)$
\begin{multline}
\frac{d}{dt} \int_{0}^{\infty} f(t,x) \phi(x) dx \\
= \frac{1}{2} \Big \langle \int_{0}^{\infty} \int_{0}^{\infty} (\phi(x_i')+\phi(x_j')-\phi(x_i)-\phi(x_j))\\
f(x_i)f(x_j)dx_i dx_j \Big \rangle.
\end{multline}
It is very useful because the choice $\phi(x)=e^{-sx}$ gives (after some calculations) the Boltzmann equation for the Laplace transform $\hat f$ of $f$
\begin{multline}
\frac{\partial \hat{f} (t, s)}{\partial t} + \hat{f} (t, s) \\
= \frac{1}{2} \Big \langle \hat{f}(t,(\lambda + (1-\lambda)\epsilon)s) \hat{f}(t,(1-\lambda)\epsilon s)
\\+ \hat{f}(t,(1-\lambda)(1-\epsilon)s) \hat{f}(t,1-(1-\lambda)\epsilon s)    \Big \rangle .
\end{multline}
For the steady state, and if $\epsilon$ is drawn randomly from a uniform distribution, the previous equation reduces to
\begin{equation}
s \hat f(s) = \frac{1}{1-\lambda} \int_{0}^{(1-\lambda)s} \hat{f}(\lambda s+y) \hat{f}(y)dy,
\label{laplace}
\end{equation}
which coincides with results of \cite{Repetowicz2005}.
The Taylor expansion of $\hat f(s)$ can be derived by substituting the expansion
$\hat{f}(s)=\sum_{p=0}^{\infty }(-1)^{p}m_{p}s^{p}$ in \eqref{laplace}. Since $\hat f(-s)$
is the moment-generating function we have $\la x^{k} \ra=m_{k}\cdot k!$.
With this method Repetowicz et al. \cite{Repetowicz2005} obtained the recursive formula
\begin{equation}
m_{p}=\sum_{q=0}^{p}m_{q}m_{p-q}\tilde{C}
_{q}^{(p)}(\lambda )
\label{recursive}
\end{equation}
with
\begin{equation}
\tilde{C}_{q}^{(p)}(\lambda )=%
\frac{\int_{0}^{(1-\lambda )}\left( \lambda +\eta \right) ^{q}\eta
^{p-q}d\eta }{1-\lambda },
\notag
\end{equation}
\begin{equation}
\tilde{C}_{q+1}^{(p)}=\frac{(1-\lambda
)^{p-q-1}-(q+1)\tilde{C}_{q}^{(p)}}{p-q},
\notag
\end{equation}
\begin{equation}
\tilde{C}_{0}^{(p)}=\frac{(1-\lambda )^{p}}{p+1}.
\end{equation}
Now with this formula one can obtain the first four moments and they match the ones found in
the previous section eqs.(14-16), which confirms that the Gamma-distribution is not the stationary distribution.

\section{Upper bound form at low wealth range}
\begin{figure}
    \begin{center}
        \includegraphics[width=\linewidth]{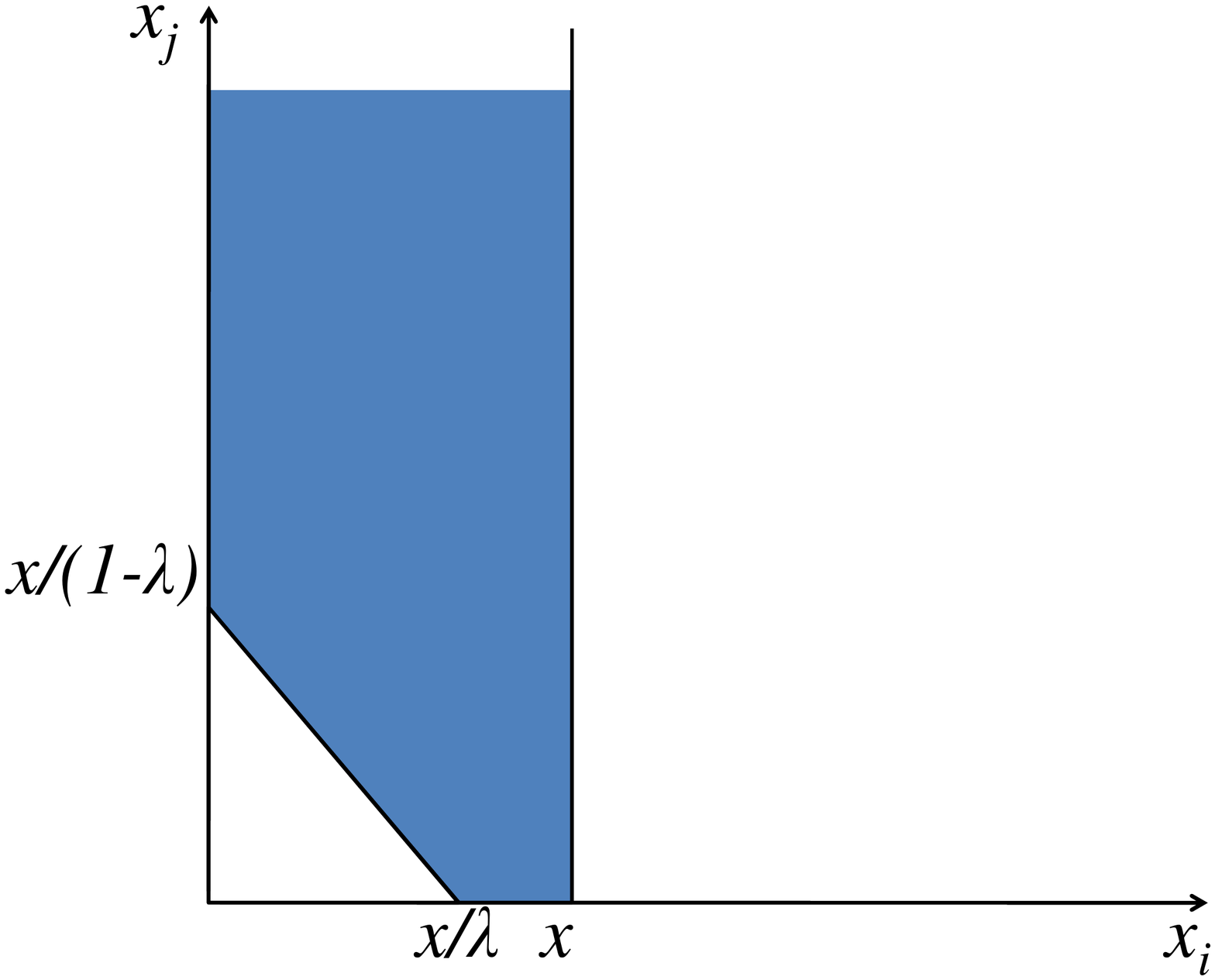}
    \end{center}
    \caption{Region of integration}
\label{fig:IntegrationDomain}
\end{figure}

From equation \eqref{distrib}
\begin{equation}
    X \stackrel{\text{d}}{=} \lambda X_i + \epsilon(1-\lambda)(X_i+X_j), \notag
\end{equation}
we have for all $x\geq 0$
\begin{equation}
   \mathbb{P}[X \leq x] = \mathbb{P}[\lambda X_i + \epsilon(1-\lambda)(X_i+X_j) \leq x],
\end{equation}
where $\mathbb{P}[.]$ means the probability of the event inside the brackets.
If the number of agents in the market is large, the distributions of different agents are independent. Then
\begin{align}
   \int_{0}^{x}{dx f(x)} &= \int_0^\infty d x_i f(x_i)  \int_0^\infty {d}x_j f(x_j)\nonumber\\
   & \times \int_0^1 {d}\epsilon \Theta \left[x-\lambda x_i + \epsilon( 1 -\lambda)( x_i + x_j)\right],
\end{align}
where $\Theta$ is the Heaviside step function. Taking the derivative with respect to $x$ in both sides, we have
\begin{align}
f(x) &= \int_0^\infty dx_i f(x_i)  \int_0^\infty dx_j f(x_j) \nonumber \\
&\times \int_0^1 d\epsilon \delta[ x - \lambda x_i - \epsilon ( 1 -\lambda) ( x_i + x_j) ].
\end{align}
This equation is an integral equation for $f(x)$. As mentioned earlier, we are not able to solve it in closed form. However, one can simplify the equation, by doing the integral over $\epsilon$. Then the $\delta$-function will contribute only if we have the following constraints
\begin{equation}
 0 \le x_i \le x/\lambda,
\end{equation}
\begin{equation}
 \frac{x - x_i}{1 -\lambda} \le x_j,
\end{equation}
\begin{equation}
 0 \le x_j.
\end{equation}
The range defined by these constraints is shown in figure \ref{fig:IntegrationDomain}. In this range, the derivative of the argument of the delta function with respect to $\epsilon$ is just $ (x_i + x_j) ( 1 - \lambda)$. And, hence
we get
\begin{equation}
f(x) = \frac{1}{1 - \lambda} \int_0^{x/\lambda} dx_i f(x_i) \int_{\max\left(\frac{x- x_i}{1 - \lambda}, 0\right)} ^{\infty} dx_j f(x_j) \frac{1}{x_i + x_j}.
\end{equation}
This immediately gives
\begin{equation}
    f(x) \le C \int_0^{x/\lambda} f(x_i) {d}x_i ,
\label{eq8}
\end{equation}
where
\begin{equation}
    C =  \frac{1}{1 -\lambda} \int_0 ^{\infty} dx_j f(x_j) \frac{1}{x_j}.
\label{ceq}
\end{equation}
We assume that $f$ decays fast enough near 0 so that the integral in \eqref{ceq} is well defined.
Now \eqref{eq8} may be rewritten by rescaling the variable, as
\begin{equation}
f( \lambda x) \le C \int_0^x dx_i f(x_i).
\label{eq9}
\end{equation}
We now use the observation that for $\lambda>0$ the numerically determined  $f(x)$ is a continuous function with a single maximum, say at $x_0$ (see \fg{fig:gamma}). Then for all $ x \le x_0$, the integrand \eqref{eq9} takes its maximum value at the right extreme point, i.e. when $x_i = x$. This then gives us
\begin{equation}
f(\lambda x) \le C x f(x), {\rm ~for} ~~x ~\le x_0.
\end{equation}
Iterating this equation, we get
\begin{equation}
f( \lambda^r x) \le  C^r \lambda^{r(r-1)/2} x^r f(x).
\end{equation}
We can set $x = x_0$ in the above equation, giving
\begin{equation}
f(\lambda^r x_0) \le  C^r \lambda^{r(r-1)/2} x_0^r f(x_0).
\end{equation}
Then taking $r\approx-\log x$ and rescaling the variables, we get
\begin{equation}
f(x) = \mathcal{O} \left( x^{\alpha} \exp[ - \beta \, (\log x)^2] \right ),
\label{upper}
\end{equation}
as $x \rightarrow 0$, where $\alpha$ and $\beta (> 0)$ are two constants dependent on $\lambda$. The Gamma-distribution decays slower than the rhs in \eqref{upper} when $x\rightarrow 0$. The expression \eqref{upper} gives an upper bound form at low wealth range and confirms again that
the distribution of the global saving propensity model is not a Gamma-distribution.

\section{Discussion and outlook}
We have used different approaches to show that the correct form of the wealth distribution cannot be the Gamma distribution. We have also derived an analytical form of an upper bound at low wealth range see Eq. \eqref{upper}.  This is an analytically calculated upper bound but the closed form of the solution to \eq{sp1} is still an open question.

As a further generalization, the agents could be assigned different saving propensities $\lambda_i$
\cite{Chatterjee2003a,Chatterjee2004a,Chatterjee2005a,Repetowicz2005,Patriarca2005a,Chakraborti2009}.
In particular, uniformly distributed $\lambda_i$ in the interval $(0,1)$ have been studied numerically in Refs.~\cite{Chatterjee2003a,Chatterjee2004a}.
This model is described by the trading rule
\begin{eqnarray}
  x_i' &=&
  \lambda_i x_i + \epsilon [ (1-\lambda_i) x_i + (1-\lambda_j) x_j ] \, ,
  \nonumber \\
  x_j' &=&
  \lambda_j x_j + (1-\epsilon) [(1-\lambda_i) x_i + (1-\lambda_j) x_j ] \, ,
  \label{sp2}
\end{eqnarray}
or, equivalently, by a $\Delta x$ (as defined in Eq.~(\ref{basic0})) given by
\begin{equation}
  \Delta x
  =  (1-\epsilon) (1-\lambda_i) x_i - \epsilon (1-\lambda_j) x_j  \, .
\end{equation}
One of the main features of this model, which is supported by theoretical considerations \cite{Mohanty,Chatterjee2005a,Repetowicz2005}, is that the wealth distribution exhibits a robust power-law at large values of $x$,
\begin{equation}\label{f-power}
  f(x) \propto x^{-\alpha - 1} \, ,
\end{equation}
with a Pareto exponent $\alpha =1$ largely independent of the details of the $\lambda$-distribution.

\begin{acknowledgments}
The authors are grateful to D. Dhar for critical reading of the manuscript and inputs in finding the functional form of the wealth distribution at low range. The authors also acknowledge F. Abergel, N. Millot, M. Patriarca, M. Politi and A. Chatterjee for critical discussions or comments, and B.K. Chakrabarti for pointing out reference \cite{Chakrabarti2010}.
AC is grateful to Department of Theoretical Physics, TIFR for the kind hospitality where part of the work was initiated.
\end{acknowledgments}


\end{document}